\documentclass[conference]{IEEEtran}

\usepackage{cite}

\usepackage[T1]{fontenc}

\usepackage[utf8]{inputenc}

\usepackage[nopatch=footnote]{microtype}

\usepackage{inconsolata}

\PassOptionsToPackage{table}{xcolor}

\usepackage[utf8]{inputenc}
\usepackage[T1]{fontenc}
\usepackage[english]{babel}

\usepackage{graphicx}
\usepackage{booktabs}
\usepackage{multirow}
\usepackage{array}
\usepackage{float}
\usepackage{adjustbox}
\usepackage{wrapfig}

\usepackage{amsmath}
\usepackage{mathtools}
\usepackage{bm}

\usepackage{algorithm}
\usepackage{algorithmicx}
\usepackage[noend]{algpseudocode}

\usepackage{xcolor}

\usepackage{enumitem}
\usepackage{microtype}
\usepackage{balance}
\usepackage{url}
\usepackage{hyperref}
\usepackage{cleveref}
\renewcommand*{\thesubsubsection}{\thesubsection~(\arabic{subsubsection})}

\usepackage{listings}
\usepackage{xspace}

\usepackage{fancyvrb}
\usepackage{comment}

\usepackage{pifont}
\usepackage{tikz}
\usepackage{circledtext}

\usepackage{booktabs}
\usepackage{multirow}

\usepackage{soul}

            \usepackage{listings}
            
\newcommand{\code}[1]{\texttt{\lstinline{#1}}}

\usepackage{tcolorbox}
\tcbuselibrary{breakable}

\lstdefinestyle{JavaStyle}{
    language=Java,
    basicstyle=\ttfamily\footnotesize,
    keywordstyle=\color{blue}\bfseries,
    commentstyle=\color{gray}\itshape,
    stringstyle=\color{red},
    numbers=left,
    numberstyle=\tiny\color{gray},
    stepnumber=1,
    showstringspaces=false,
    breaklines=true,
    frame=none,
    captionpos=b
}

\usepackage{amssymb}

\newcommand{\spec}{\mathbf{Spec}}
\newcommand{\specprobe}{\textsc{SpecProbe}}

\newcommand{\synC}{\text{\normalfont Exec@k}}
\newcommand{\semC}{\text{\normalfont Sem@k}}

\newcommand{\bugD}{\text{\normalfont BDR}}

\newtcolorbox{findingbox}[1][]{colback=black!5, colframe=black!40, boxrule=0.4pt, arc=1pt, left=4pt, right=4pt, top=2pt, bottom=2pt, before upper={\textbf{\small #1:}\enspace}, before skip=3pt, after skip=3pt}

\title{SpecProbe: Probing Intent--Implementation Misalignments with Generated Executable Postconditions}

\author{
\IEEEauthorblockN{Gehao Zhang}
\IEEEauthorblockA{University of Massachusetts Amherst\\
gehaozhang@umass.edu}
\and
\IEEEauthorblockN{Zhenting Wang}
\IEEEauthorblockA{Rutgers University\\
zhenting.wang@rutgers.edu}
\and
\IEEEauthorblockN{Juan Zhai}
\IEEEauthorblockA{University of Massachusetts Amherst\\
juanzhai@umass.edu}
}

\begin{document}
\maketitle

\begin{abstract}
  Ensuring that code faithfully reflects its documented intent is a fundamental challenge in software engineering, but the informal natural language specifications in real repositories cannot be automatically compiled into executable checks and run against implementations.
  We introduce \emph{behavioral probes}: executable postconditions generated from natural language documentation, injected into programs, and checked during test execution to reveal where documented intent and implementation behavior align or diverge.
  We present \specprobe{}, a repository-aware pipeline that generates behavioral probes and classifies each as non-executable, matched, or mismatched.
  A structured four-field schema (\texttt{import}, \texttt{pre-ghost}, \texttt{post-ghost}, \texttt{condition}) makes generated postconditions injectable and compilable in real repositories; reasoning supervised fine-tuning on ${\sim}$1.5K automatically curated examples enables a compact 7B model to generate probes without continued API access.
  On Defects4J, \specprobe{}'s 7B model achieves GPT-4o-comparable semantic consistency (Sem@1) at 141$\times$ lower inference cost.
  The pipeline reveals 80 of 726 historical bugs, and for 95\% of them the original failing-test stack trace omits the method checked by the probe.
  On Python icontract generation, \specprobe{} surpasses GPT-5-mini and GPT-o3, demonstrating transfer to a different language and contract format.
  These results show that behavioral probes complement failing-test stack traces with diagnostic evidence grounded in the selected method's documented intent.
\end{abstract}


\section{Introduction}\label{sec:intro}

In the emerging agentic era of software engineering, developers increasingly use natural language~(NL) to interact with LLM-based agents that generate, modify, and maintain code~\cite{jimenez2024swe,yang2024swe}.
This makes the alignment between human intent and code behavior both more pervasive and more critical: when the primary interface is NL, ensuring that the resulting code faithfully reflects the stated intent becomes a first-class concern.
Studies show that documentation-code inconsistencies are widespread: iComment~\cite{tan2007icomment} found that comment-code inconsistencies account for a substantial fraction of OS kernel bugs, @TComment~\cite{tan2012tcomment} identified systematic Javadoc violations in production Java projects, and Zhou et al.~\cite{zhou2017analyzing} detected directive defects across hundreds of real API methods.
NL descriptions in Javadoc, docstrings, and API documentation can conflict with the implementations they describe, and developers have no automated way to detect these misalignments at scale.

The root cause is a fundamental representation gap: NL documentation is informal and human-readable, while automated behavioral checking requires precise, executable conditions that can be compiled and run against the implementation.
The natural response is to generate postconditions automatically from NL documentation.
Prior pattern- and search-based methods~\cite{tan2007icomment,tan2011acomment,zhai2020c2s,goffi2016automatic,blasi2018translating} require heavy manual effort and lack generalizability.
Recent LLM-based approaches~\cite{endres2024can,xie2025effective} show that postconditions can be generated from NL and checked through test execution.
However, condition-only outputs provide limited support for repository dependencies and properties that compare pre-call and post-call state.
We study how to make these generated conditions executable in repository context and interpret their test outcomes as evidence of alignment or disagreement.

We propose to close this gap by generating \emph{behavioral probes}: executable postcondition candidates that can be injected into source code and checked during test execution to triangulate NL intent, generated contracts, and observed implementation behavior.
If a probe compiles, executes, and does not violate during testing, it provides evidence that the documented intent and the implementation are \emph{aligned}.
If it violates, it reveals a \emph{misalignment} among three artifacts (the NL description, the generated postcondition, and the code).
This framing turns generated postconditions into concrete diagnostic signals: a matched probe provides evidence of alignment, while a mismatched probe exposes a specific semantic disagreement at the selected method for developer triage.

We present \specprobe{}, a repository-aware pipeline with three core components.
A \emph{prompt builder} combines the method's NL documentation with class- and repository-level code context, providing the model with the semantic, structural, and dependency information necessary to generate precise executable postconditions.
A \emph{structured specification schema} with four fields (\texttt{import}, \texttt{pre-ghost}, \texttt{post-ghost}, \texttt{condition}) bridges generated text and executable artifacts, making repository dependencies and pre-/post-state references explicit for deterministic injection into source code.
An \emph{instrumentation-based probe checker} injects each candidate into source code and runs existing tests to classify it as non-executable, matched, or mismatched.
To keep inference cost low, we apply reasoning supervised fine-tuning~(SFT) to Qwen2.5-Coder-7B~\cite{hui2024qwen2} using ${\sim}$1.5K automatically curated examples.
The resulting 7B model runs without frontier API access, amortizing the one-time curation cost across all future inference.

We evaluate \specprobe{} on Java Defects4J~\cite{just2014defects4j} for alignment assessment and bug-revealing, and on Python HumanEval~\cite{liu2023your} to assess transfer to a different language and contract format.
On Defects4J, \specprobe{}'s 7B model achieves GPT-4o-comparable semantic consistency (43.0\% vs.\ 45.3\%) at 141$\times$ lower inference cost and generates at least one correct probe for over 80\% of target methods.
Among open-source models, \specprobe{} ranks first on all semantic consistency metrics despite being the smallest model in the comparison, outperforming models with up to 4.6$\times$ as many parameters.
\specprobe{} reveals 80 of 726 historical bugs in methods selected from the fixes. The original failing-test stack trace omits the target method in 95\% of these cases and its class in 91\%.
On Python icontract generation, \specprobe{} surpasses GPT-5-mini and GPT-o3 on both semantic consistency and mutant-distinguishing rate, showing that the pipeline transfers to Python/icontract and remains effective with a different curator.

Our contributions are:
\begin{itemize}[leftmargin=*]
    \item \textbf{Executable specification schema~(\S\ref{sec:design}).} Our primary technical contribution is making generated specifications directly usable as executable program artifacts. The four-field schema explicitly represents dependencies, pre-call state capture, post-call computation, and the final condition, allowing generated specifications to be instrumented into program executions for testing and behavioral checking.
    \item \textbf{Efficient adaptation for probe generation~(\S\ref{sec:rq2}).} We show that this capability can be adapted to a 7B model using approximately 1.5K automatically curated, task-specific examples with reasoning SFT, avoiding reliance on frontier models at inference time.
    \item \textbf{Diagnostic evidence beyond stack traces~(\S\ref{sec:results}).} Bug-revealing probes provide diagnostic evidence beyond the failing-test stack trace once a method has been selected for investigation. For 95\% of bugs revealed by \specprobe{}, the target method is absent from the original failing-test stack trace. The probe instead exposes the violated behavioral condition at the selected method.
\end{itemize}

\section{Background}\label{sec:bg}

\noindent\textbf{Postconditions and behavioral probes.}
A \emph{postcondition} is a logical predicate that must hold after a method execution~\cite{meyer1992contract,leavens1999jml}.
When a postcondition can be compiled, injected into source code, and evaluated at runtime, it becomes an \emph{executable specification}, i.e., a machine-checkable behavioral contract.
We use the term \emph{behavioral probe} for an executable postcondition candidate generated from NL documentation and used to assess intent--implementation alignment.
A probe is not a ground-truth oracle: passing during test execution provides evidence of alignment, whereas a violation reveals a disagreement among the NL description, the postcondition, and observed behavior.

\noindent\textbf{Ghost variables.}
Postconditions that express state-change properties must compare pre-call and post-call values.
A \emph{ghost variable}, introduced solely for specification purposes~\cite{leavens1999jml}, snapshots a value before or after the method call so the postcondition can reference it.
For example, to assert that a parsing cursor advances, a \texttt{pre-ghost} variable records the cursor position before the call, and the postcondition checks that the post-call position exceeds it.
Ghost variables exist only in the instrumentation wrapper, not in the program logic.

\noindent\textbf{Reasoning SFT.}
SFT adapts a pretrained model by training on input--output pairs; \emph{reasoning SFT} also includes intermediate reasoning traces in the training target, encouraging the model to decompose the task, track dependencies, and self-correct before committing to an output.
In our setting, each training instance comprises a prompt (NL documentation and repository context), a reasoning trace, and a structured executable postcondition.

\section{Motivating Example}\label{sec:motivating-example}

\begin{figure}[tb]
    \centering
    \footnotesize
    \includegraphics[
    width=0.45\textwidth,
    trim=0cm 0.2cm 0cm 0cm,clip
    ]{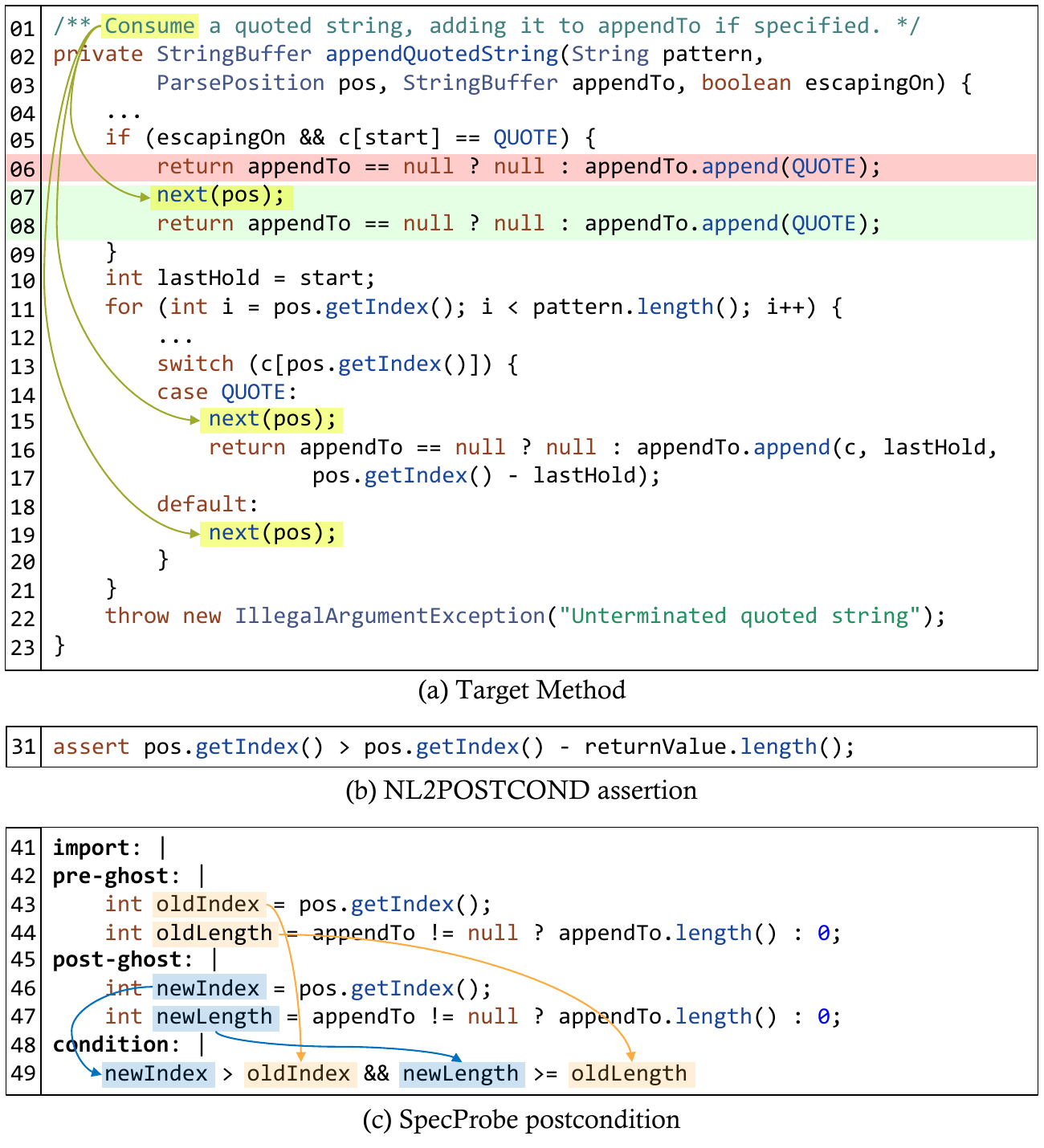}
    \caption{Motivating Example.
    }
    \label{fig:moti}
\end{figure}

\noindent\textbf{Structured schema enables executable state-change checking.}
\autoref{fig:moti}(a) shows an example from the \code{Lang}~\cite{apache-commons-lang-github} project.
The target method \code{appendQuotedString} is documented to ``consume a quoted string,'' implying that the parsing position \code{pos} must advance, which is done by calling \code{next(pos)} as it reads characters.
However, the buggy version omits \code{next(pos)} in the escaped-quote branch (line~6), leaving \code{pos} unchanged and violating the documented ``consume'' semantics, although the returned buffer appears correct.
The fix inserts \code{next(pos)} at line~7 before returning.
Such bugs are hard to catch by output inspection alone, yet a documentation-derived executable postcondition (checking that \code{pos} advances) exposes them automatically.

\autoref{fig:moti}(b) shows the assertion generated by NL2POSTCOND~\cite{endres2024can} for the same method (line~31): \code{pos.getIndex() > pos.getIndex() - returnValue.length()}, a single assertion injected after the end of the target method.
It is problematic in two ways.
First, if \code{returnValue} is null, evaluating \code{returnValue.length()} throws a \texttt{NullPointerException} before the assertion is checked.
Second, NL2POSTCOND~\cite{endres2024can} generates a plan before each postcondition. The generated assertion (line~31) can not implement its plan: ``Checks that the index of the ParsePosition has increased after the method execution.'' It cannot read the pre-state; it reads the post-call index on both sides.
When \code{returnValue} is non-null, it only checks that the returned buffer is nonempty, so even when it holds, it does not establish that \code{pos} advanced.

\autoref{fig:moti}(c) shows the postcondition generated by \specprobe{}.
Our schema lets the model introduce ghost variables (\S\ref{sec:bg}): \texttt{pre-ghost} variables (lines~43--44) record values before the call, and \texttt{post-ghost} variables (lines~46--47) capture values after it.
Here, the model declares \code{oldIndex = pos.getIndex()} as a pre-ghost (line~43) and \code{newIndex = pos.getIndex()} as a post-ghost (line~46), then checks \code{newIndex > oldIndex} at line~49, directly formalizing the documented ``consume'' semantics.
It also records the length of \code{appendTo} before and after the call (lines~44, 47) to ensure that the output buffer does not shrink.
Executed against the buggy version, this postcondition is violated and the bug is caught automatically.
Because the schema exposes the pre-call snapshot to the condition, it can express this state change, which a condition-only format cannot.

\smallskip
\noindent\textbf{Reasoning enables self-correction.}
A second gap is the lack of intermediate reasoning: without it, a model commits to its first documentation interpretation even when it is incomplete.
Consider \code{generateToolTipFragment(String toolTipText)} from \code{Chart}~\cite{jfree-jfreechart-github}, which produces an HTML tooltip fragment for an image map tag.
NL2POSTCOND generates:

\noindent
{\ttfamily\footnotesize
\hspace*{1.3em}ret != \textcolor[HTML]{954121}{null} \&\& ret.\textcolor[HTML]{19469D}{contains}(toolTipText)}

This condition ignores HTML escaping, so raw unescaped text can wrongly satisfy it.
With reasoning SFT, \specprobe{}'s trace first considers the same flawed check, then recognizes that, e.g., \code{"Hello <World>"} should appear as \code{"Hello \&lt;World\&gt;"}, and revises the specification by computing the escaped string in a pre-state ghost variable:

\noindent
{\ttfamily\footnotesize
\hspace*{1.3em}\textcolor[HTML]{408080}{// pre-state ghost variable}\\
\hspace*{1.3em}\textcolor[HTML]{445588}{String} escapedToolTip = toolTipText.\textcolor[HTML]{19469D}{replace}(\textcolor[HTML]{219161}{"\&"}, \textcolor[HTML]{219161}{"\&amp;"})\\
\hspace*{3.3em}.\textcolor[HTML]{19469D}{replace}(\textcolor[HTML]{219161}{"<"}, \textcolor[HTML]{219161}{"\&lt;"}).\textcolor[HTML]{19469D}{replace}(\textcolor[HTML]{219161}{">"}, \textcolor[HTML]{219161}{"\&gt;"});\\
\hspace*{1.3em}\textcolor[HTML]{408080}{// postcondition}\\[-4pt]
\hspace*{1.3em}ret != \textcolor[HTML]{954121}{null} \&\& ret.\textcolor[HTML]{19469D}{contains}(escapedToolTip)}

This self-correction yields a postcondition that rejects unescaped HTML output. Without human annotation, the reasoning model inferred the requirement from documentation alone.

\begin{figure*}[t]
    \centering
    \footnotesize
    \includegraphics[
    width=.9\textwidth,
    ]{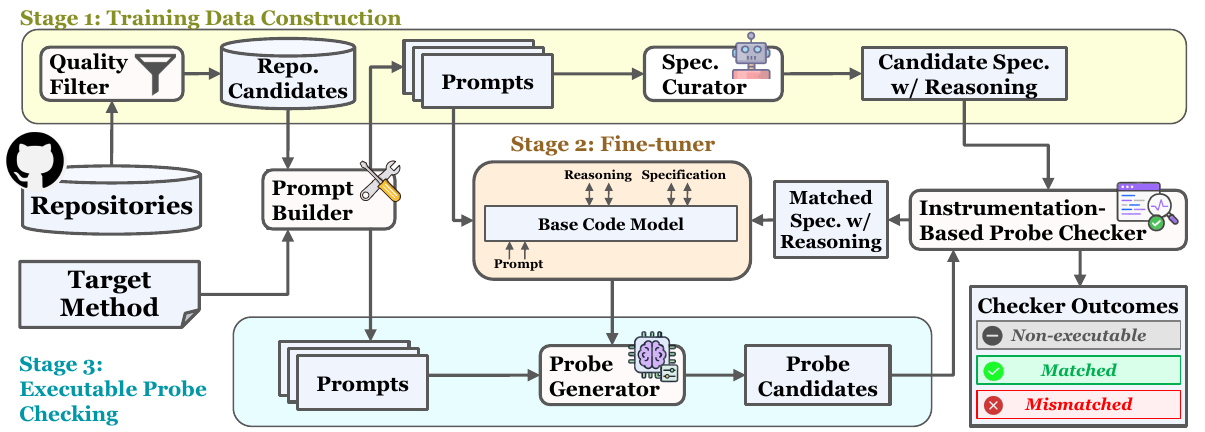}
\caption{\specprobe{} Pipeline: Data Construction, Reasoning Supervised Fine-Tuning, and Probe Checking.
}\label{fig:overview}
\end{figure*}

\section{Methodology}\label{sec:design}

\subsection{Overview}
\autoref{fig:overview} illustrates the \specprobe{} pipeline.
In \textbf{Training Data Construction}, a quality filter selects well-documented, well-tested methods from GitHub; the Prompt Builder assembles NL documentation with repository context; the Specification Curator (a frontier reasoning model) produces reasoning traces and structured postconditions; and the probe checker retains only matched candidates for training (\S\ref{sec:probe-checker}).
In \textbf{Reasoning SFT}, a compact 7B model is fine-tuned on these prompt--reasoning--postcondition triples, learning to decompose the task and justify each decision before emitting the final executable output.
In \textbf{Probe Checking}, the fine-tuned model generates candidates for new target methods, which the probe checker injects into source code and runs against existing tests, labeling each non-executable, matched, or mismatched; a mismatch signals disagreement among NL documentation, the generated postcondition, and observed behavior.

\subsection{Quality Filter for Repository Selection}
\specprobe{} curates training data from public GitHub repositories with a two-level quality filter.
At the repository level, we retain projects with high popularity (star and fork counts), active maintenance (recent commits and issue resolutions), and comprehensive documentation, as these indicators correlate with well-structured code and reliable test suites.
At the method level, each candidate must
(1)~have an NL description from which the Prompt Builder can extract behavioral intent, and
(2)~be exercised by at least one test, since the probe checker relies on test execution.
Given the abundance of eligible methods, we randomly select among them to ensure diversity and coverage across different projects.

\subsection{Prompt Builder}\label{sec:promptBuilder}

\begin{figure}
    \centering
    \footnotesize
    \includegraphics[
    width=.85\linewidth,
    trim=0cm 0cm 0cm 0cm,clip
    ]{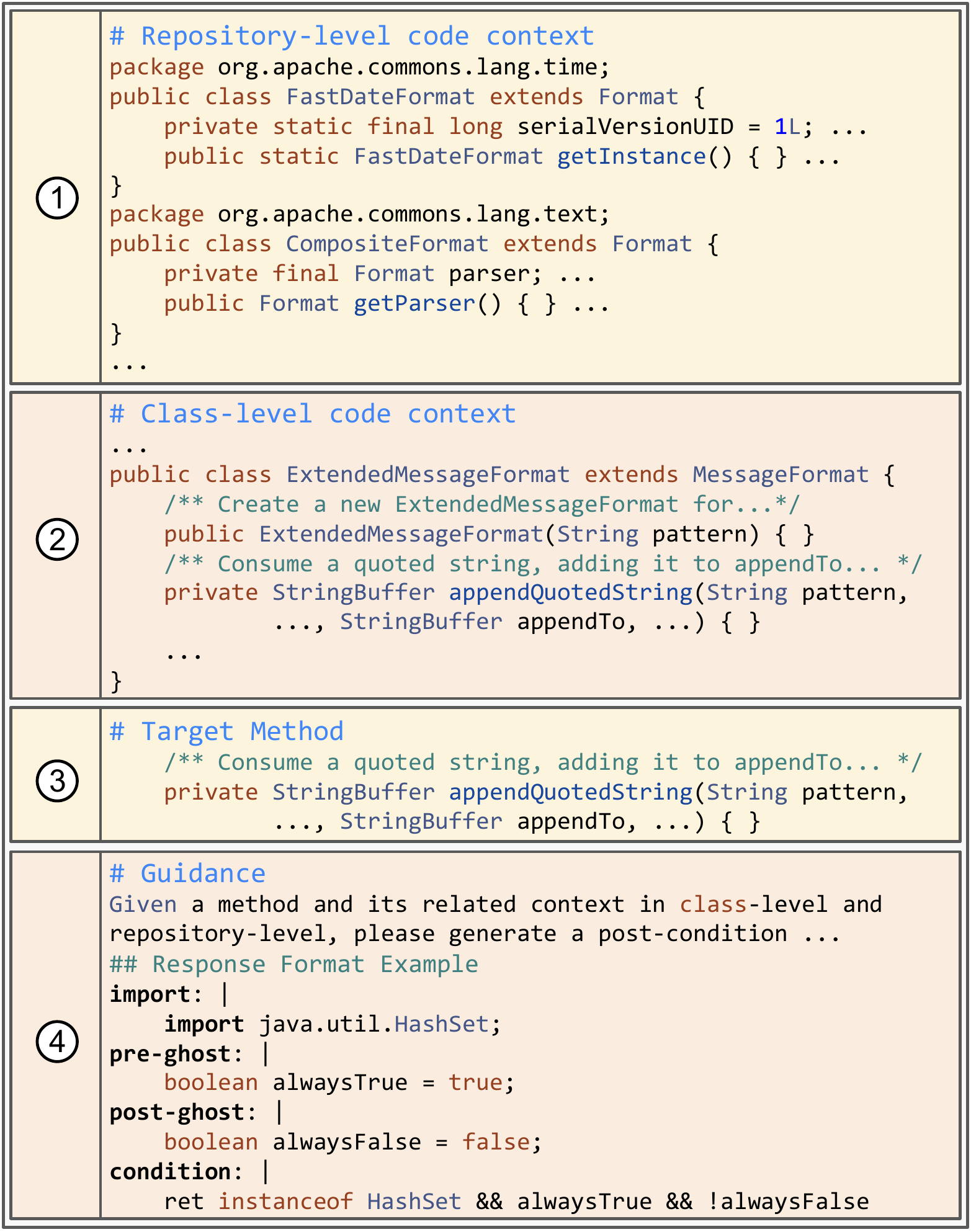}
    \caption{LLM Prompt Example.
    }\label{fig:prompt}
\end{figure}

\specprobe{} automatically constructs structured prompts (\autoref{fig:prompt}) with four components: (\textcircled{\small{1}})~repository-level context, (\textcircled{\small{2}})~class-level context, (\textcircled{\small{3}})~target method signature with its NL description, and (\textcircled{\small{4}})~task-specific instructions.
The \textit{Prompt Builder} produces fine-tuning inputs during training data construction and dynamically builds prompts for input methods at inference.

The repository- and class-level contexts~(\textcircled{\small{1}} and \textcircled{\small{2}}) supply related class skeletons with the cross-class dependencies needed to interpret method behavior and infer accurate postconditions.

The instruction component~(\textcircled{\small{4}}) defines the task and enforces our four-field output schema.
\noindent
\texttt{import} declares package dependencies required by the postcondition.
\texttt{pre-ghost} snapshots variables before the method call so their original values can be referenced in the condition.
\texttt{post-ghost} captures values computed after the call (e.g., the return value) for use in the condition.
\texttt{condition} is the executable assertion over pre-ghost, post-ghost, and program state.
Together, they bridge generated text and executable artifacts, enabling state-change properties and non-local dependencies that prior condition-only formats~\cite{endres2024can} cannot express.

\subsection{Specification Curator}\label{sec:specificationcreator}

\begin{figure}[!b]
    \centering
    \footnotesize
    \includegraphics[width=.98\linewidth,
    trim=0cm 0.2cm 0cm 0cm,clip
    ]{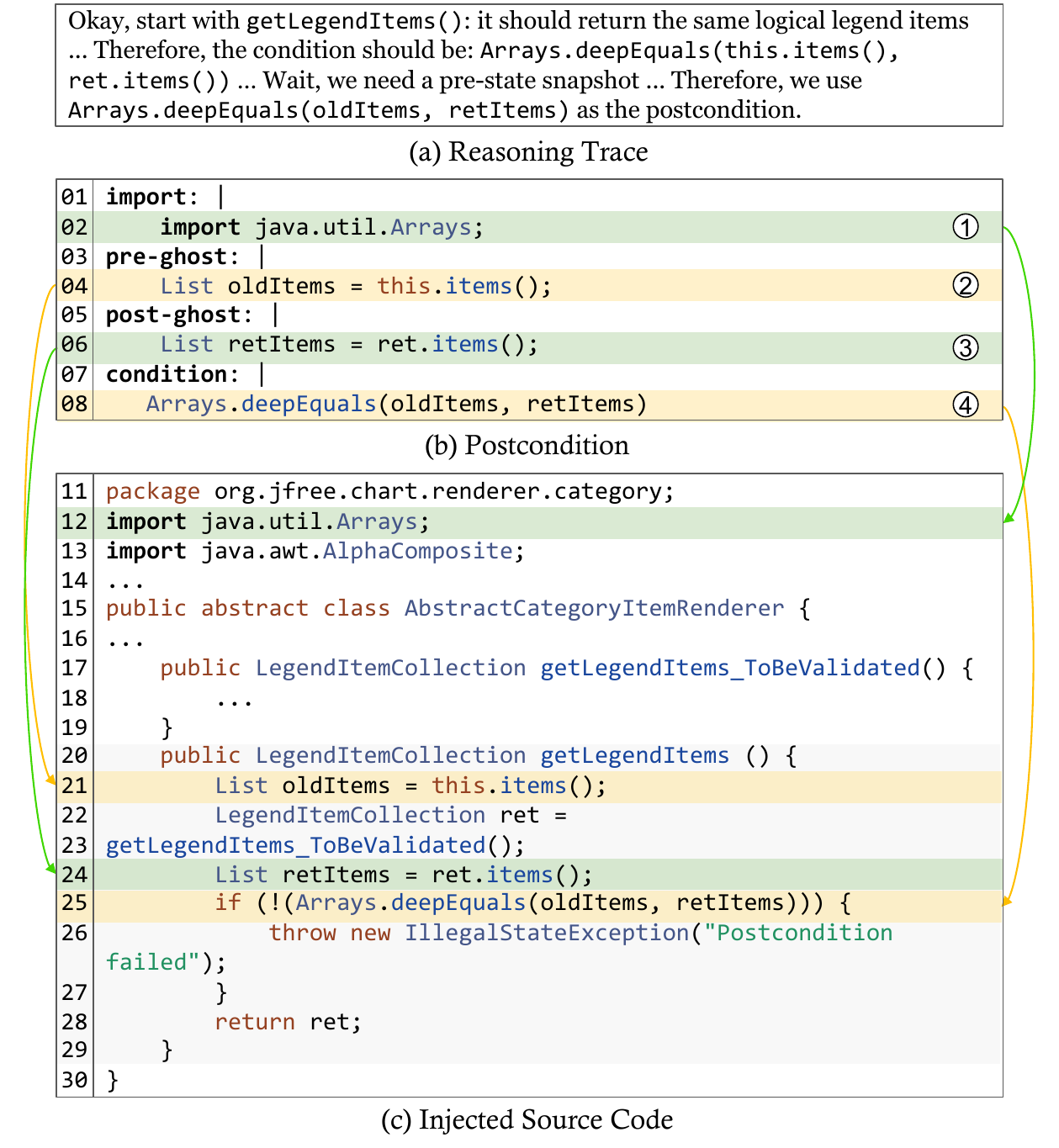}
    \caption{Reasoning, Postcondition, and Injection Example.}
    \label{fig:output_example}
\end{figure}

Our curator invokes a frontier reasoning model for schema-conforming candidate postconditions with reasoning traces.

\autoref{fig:output_example}(b) shows an example for \texttt{getLegendItems()}.
The postcondition \texttt{Arrays.deepEquals(oldItems, retItems)} (lines~7--8) calls \texttt{deepEquals} from \texttt{Arrays}, which requires an explicit dependency declaration \texttt{import java.util.Arrays;} in the \texttt{import} block (lines~1--2).
Its arguments are ghost variables:
\texttt{oldItems} snapshots the receiver’s pre-state in \texttt{pre-ghost} (line~4),
and \texttt{retItems} captures the post-state from the return value in \texttt{post-ghost} (lines~5--6).

Each candidate's reasoning trace (\autoref{fig:output_example}(a)) records step-by-step deductions from NL documentation to the postcondition, serving as intermediate supervision in fine-tuning.

\subsection{Instrumentation-Based Probe Checker}\label{sec:probe-checker}
The probe checker includes test execution. It filters training examples in training and classifies probes in inference.

\noindent\textbf{Instrumentation.}
The checker injects each candidate into the target method's source file and runs all exercising tests; instrumenting the source rather than test files ensures that every test reaching the method exercises the postcondition without requiring knowledge of test structure.

As \autoref{fig:output_example}(c) illustrates, we rename the target method by appending \texttt{\_ToBeValidated} and create a wrapper with the original signature, which invokes the renamed method and stores its return value (if any) in \texttt{ret} (lines~22--23).
We (i) insert the required \texttt{import} statements (line~12), (ii) emit \texttt{pre-ghost} statements before the invocation to snapshot pre-state values (line~21), (iii) emit \texttt{post-ghost} statements after it to compute post-state values, possibly derived from \texttt{ret} (line~24), and (iv) evaluate \texttt{condition} before returning (line~25).
If the condition is violated, the wrapper throws an exception (e.g., \texttt{IllegalStateException}, line~26), causing the test to fail.

\noindent\textbf{Checker outcomes.}
The checker reports one of three outcomes:
(1)~\textit{Non-executable} candidates for which parsing, injection, compilation, formatting, dependency resolution, or test execution fails, including exceptions other than the checker's explicit postcondition-violation exception (e.g., an exception thrown before the target method returns);
(2)~\textit{Matched} candidates that can be parsed, injected, compiled, and executed without postcondition violation under the tests; and
(3)~\textit{Mismatched} candidates for which the target method returns normally and the injected postcondition explicitly reports a violation during test execution.

\noindent\textbf{Training-data filtering.}
During training-data construction, only candidates matched against the reference implementation are retained as supervision examples; non-executable and mismatched candidates are discarded, as neither can serve as reliable positive training signal.

\noindent\textbf{Inference-time interpretation.}
At inference time, non-executable candidates are discarded by default as they provide no semantic evidence.
Matched and mismatched candidates are behavioral probes:
matched probes agree with test-observed behavior and can be reviewed as candidate contracts or regression checks;
mismatched probes are discrepancy reports whose violated condition and failing test give developers a concrete triage point to decide whether the implementation, postcondition, or NL description should change.
Probes are stored as persistent YAML artifacts whose task metadata links each to its target method; they can be inspected, rerun, and optionally version-controlled alongside the code.
Probes need not be regenerated for every code version: a saved probe remains reusable while the target method's informal specification applies, and can be regenerated when that specification or relevant context changes.

\noindent\textbf{Optional revision of non-executable probes.}
\specprobe{} optionally repairs non-executable probes using execution feedback: the model receives the probe and its error message and is instructed to preserve the intended postcondition while fixing compilation, type/API, or formatting errors. Repair is intentionally restricted to non-executable probes: execution errors are actionable feedback, whereas repairing a probe based on a semantic mismatch could mask a genuine documentation--implementation discrepancy that should be surfaced for triage.

\subsection{Fine-tuner and Probe Generator}\label{sec:fine-tuning}

We fine-tune a compact 7B model via reasoning SFT to generate executable probe candidates from NL prompts enriched with repository context, transferring a frontier model's probe-generation capability into a lightweight model that reasons explicitly at a fraction of the inference cost.
The resulting fine-tuned model serves as the \textit{Probe Generator} at inference time, producing probe candidates that are subsequently classified by the Probe Checker.

Each training instance comprises a prompt, reasoning logic, and an expected postcondition.
The prompt integrates repository- and class-level context, the target method, and a task-specific instruction guiding postcondition generation.
The training target concatenates the reasoning trace and structured postcondition, so the loss supervises both intermediate behavioral rationale and the final executable output.

\section{Experimental Setup}\label{sec:setup}

We design our evaluation to answer four research questions:
\begin{itemize}[leftmargin=*]
    \item \textbf{RQ1:} How does a 7B model compare with larger models on probe generation?
    \item \textbf{RQ2:} Does reasoning justify its cost?
    \item \textbf{RQ3:} What diagnostic evidence do probes provide beyond test stack traces?
    \item \textbf{RQ4:} Does the pipeline generalize across language, contract format, and curator model?
\end{itemize}

\subsection{Benchmark}

Following prior work on postcondition generation~\cite{endres2024can,ma2024specgen}, we evaluate on Defects4J v2.0.0~\cite{just2014defects4j}, a widely used benchmark spanning 17 open-source Java projects (e.g., Apache Commons Math~\cite{apache-commons-math-github}, JFreeChart~\cite{jfree-jfreechart-github}).
Projects range from 1 to 96 KLOC (median 14 KLOC), with 0.1K--8.4K methods and 0.1K--15.3K tests each.
Defects4J suits our evaluation: its paired fixed/buggy versions enable measuring bug-revealing utility, its methods have the documented NL descriptions and mature test suites our pipeline requires, and NL2POSTCOND~\cite{endres2024can} uses it as well.
Of its 835 bugs, we evaluate 726 whose patched method keeps its signature across buggy and fixed versions, has NL documentation, and is exercised by at least one test.
These yield 1,134 generation tasks; each corresponds to a method modified by the fix.

\subsection{Baselines and Evaluation Protocol}

We compare against seven baselines spanning prior work and frontier models:
GPT-5.5~\cite{openai2026gpt55},
GPT-4o~\cite{hurst2024gpt},
GPT-4 and StarChat-Alpha~\cite{li2023starcoder} (from NL2POSTCOND~\cite{endres2024can}),
Qwen2.5-Coder-7B/32B-Instruct~\cite{hui2024qwen2}, and
DeepSeek-R1-Distill-32B~\cite{guo2025deepseek}.
All non-asterisked models in \autoref{tab:model_performance} use the same 1,134 target methods.
We independently sample 10 candidates per task (temperature $= 1$, top-$p = 1$, no reranking or repetition penalties), yielding 11,340 postconditions per model, classified solely by the probe checker and no post-hoc reranking.
The four-field evaluation uses a 22,000-token output budget wherever supported; GPT-4o uses its API maximum of 16,384 tokens, and its longest completion is 508 tokens.
Post-processing strips Markdown formatting.

\subsection{Implementation Details}

\noindent\textbf{Training data curation.}
DeepSeek-R1~\cite{guo2025deepseek} serves as the specification curator, producing ${\sim}$1.5K prompt--reasoning--postcondition triples after probe-checker filtering.
Its reasoning traces are kept unchanged, without editing or post-processing; the only filtering is at the example level, where the probe checker accepts or rejects a whole triple based on whether its postcondition passes all exercising tests.

\noindent\textbf{Fine-tuning.}
We fine-tune Qwen2.5-Coder-7B-Instruct with AdamW ($\beta_1 = 0.9$, $\beta_2 = 0.95$, weight decay $1{\times}10^{-4}$), learning rate $1{\times}10^{-5}$, a cosine schedule, and 5\% warmup.

\subsection{Metrics}\label{sec:metrics}

We report three metrics under a top-$k$ protocol: \textit{Executability} ($\synC$), \textit{Test-Observed Semantic Consistency} ($\semC$), and \textit{Bug-Revealing Rate} ($\bugD$).
All three count a task as successful if at least one of the top-$k$ candidates satisfies the criterion.

\noindent\(\bullet\)\hspace{3pt}\textbf{Executability (\(\synC\))}
measures whether a top-$k$ candidate is parseable, injectable, and compilable, i.e., syntactically and structurally valid; it need not pass any test.
\begin{equation*}
    \synC = \frac{1}{N} \sum_{j=1}^{N} \mathbf{1} \left( \bigvee_{i=1}^{k} \text{Executable}(\text{Candidate}_{i,j}) \right),
\end{equation*}
where \( \text{Candidate}_{i,j} \) is the \( i \)-th candidate for the \( j \)-th task, and \(N\) is the total number of tasks.

\noindent\(\bullet\)\hspace{3pt}\textbf{Test-Observed Semantic Consistency (\(\semC\))}
measures whether a top-$k$ candidate is executable \emph{and} consistent with the fixed implementation under test execution, reflecting test-observed behavior rather than full semantic correctness.
\begin{equation*}
    \semC = \frac{1}{N} \sum_{j=1}^{N} \mathbf{1} \left( \bigvee_{i=1}^{k} \text{Consistent}(\text{Candidate}_{i,j}) \right),
\end{equation*}
where \( \text{Consistent}(\text{Candidate}_{i,j}) \) is true when it compiles, executes, and does not violate on the fixed implementation.

\noindent\(\bullet\)\hspace{3pt}\textbf{Bug-Revealing Rate (\(\bugD\))}
measures the discriminative power of generated probes: whether at least one candidate passes on the fixed version but triggers a postcondition violation on the buggy version, where the failure must be attributable to the injected postcondition rather than unrelated build or test artifacts.
Let \( \{ \spec_1, \dots, \spec_m \} \) be the postconditions generated (10 per target method) for the methods affected by a bug; the bug is revealed (counted once) if any satisfies:
\begin{equation*}
    d_i \iff \text{Pass}(\mathcal{T}, C^{\text{fixed}}_{\spec_i}) \;\land\; \text{Violate}(\mathcal{T}, C^{\text{buggy}}_{\spec_i})
\end{equation*}
\begin{equation*}
    \bugD = \frac{1}{N} \sum_{j=1}^{N} \mathbf{1} \left( \bigvee_{i=1}^{m} d_i  \right),
\end{equation*}
where \( N \) is the total number of evaluated bugs,
\( \text{Pass}(\mathcal{T}, C^{\text{fixed}}_{\spec_i}) \) denotes that the test suite passes on the fixed version instrumented with \( \spec_i \), and
\( \text{Violate}(\mathcal{T}, C^{\text{buggy}}_{\spec_i}) \) denotes that the buggy version fails due to a postcondition violation.

\section{Results}
\label{sec:results}

\begin{table*}[tb]
    \definecolor{fstc}{HTML}{AAAAFF}
    \definecolor{scdc}{HTML}{C9C9FF}
    \definecolor{trdc}{HTML}{EEEEFF}
    \newcommand{\fst}[1]{\textbf{#1}}
    \newcommand{\scd}[1]{#1}
    \newcommand{\trd}[1]{#1}
    \centering
    \footnotesize
    \caption{Probe Generation Performance (bold: best per group). The starred GPT-4 and StarChat rows report published NL2POSTCOND results~\cite{endres2024can} on 840 tasks and 525 bugs. The ``GPT-4o w.~\cite{endres2024can}'' row reports our reconstruction of NL2POSTCOND on our 1,134 tasks, following NL2POSTCOND's prompt. Both the published results and the reconstruction use NL2POSTCOND's condition-only format. All other rows use \specprobe{}'s four-field schema. Cost is USD per generated postcondition.}\label{tab:model_performance}
    \begin{tabular}{l|cc|rrrrrr|rr}
        \toprule
    Model & Size & Cost (USD) & Sem@1 & Sem@5 & Sem@10 & Exec@1 & Exec@5 & Exec@10 & BDR & \#Bugs \\ \midrule
    GPT-5.5 & \(\gtrsim\)200B & 0.117 & \textbf{71.7\%} & \textbf{84.7\%} & \textbf{88.4\%} & \textbf{95.8\%} & 97.5\% & 97.6\% & \textbf{32.1\%} & \textbf{233\textsubscript{/726}} \\
    GPT-4o & \(\gtrsim\)200B & 0.023 & 45.3\% & 77.8\% & 87.1\% & 85.4\% & \textbf{97.7\%} & \textbf{98.5\%} & 12.1\% & 88\textsubscript{/726} \\
    GPT-4o w.~\cite{endres2024can} & \(\gtrsim\)200B & 0.003 & 46.5\% & 67.6\% & 74.3\% & 70.1\% & 83.4\% & 86.7\% & 7.9\% & 57\textsubscript{/726} \\
    \midrule
    Qwen2.5-Coder & 7B & \(0.090{\times}10^{-3}\) & 12.5\% & 42.3\% & 60.8\% & 34.9\% & 80.5\% & 91.2\% & 5.0\% & 36\textsubscript{/726} \\
    Qwen2.5-Coder & 32B & \(0.441{\times}10^{-3}\) & \trd{32.6\%} & \trd{63.4\%} & \trd{73.5\%} & \fst{74.7\%} & \scd{94.5\%} & \scd{96.0\%} & \trd{10.6\%} & \trd{77\textsubscript{/726}} \\
    DeepSeek-R1-Distill & 32B & \(1.417{\times}10^{-3}\) & \scd{39.8\%} & \scd{71.7\%} & \scd{80.2\%} & \trd{67.5\%} & \trd{94.2\%} & \scd{96.0\%} & \fst{12.7\%} & \fst{92\textsubscript{/726}} \\
    \specprobe{} & 7B & \(0.165{\times}10^{-3}\) & \fst{43.0\%} & \fst{72.3\%} & \fst{80.5\%} & \scd{72.3\%} & \fst{94.7\%} & \fst{96.6\%} & \scd{11.0\%} & \scd{80\textsubscript{/726}} \\
    \midrule
    \multicolumn{11}{l}{\emph{Published NL2POSTCOND results (840 tasks, 525 bugs)}} \\
    GPT-4* & \(\gtrsim\)200B & 0.036 & 32.0\% & 57.0\% & 66.0\% & 65.0\% & 86.0\% & 89.0\% & 6.7\% & 35\textsubscript{/525}\\
    StarChat* & 16B & \(0.251{\times}10^{-3}\) & 11.0\% & 38.0\% & 55.0\% & 25.0\% & 68.0\% & 83.0\% & 3.6\% & 19\textsubscript{/525} \\
    \bottomrule
    \end{tabular}
\end{table*}

\subsection{RQ1: How does a 7B model compare with larger models on probe generation?}

\autoref{tab:model_performance} compares \specprobe{} against three proprietary models (GPT-5.5, GPT-4o, GPT-4) and four open-source baselines (Qwen2.5-Coder-7B/32B, StarChat, DeepSeek-R1-Distill-32B).
Cost is the average USD inference cost per generated postcondition~\cite{litellm2026modelpricing,openai2026apipricing}.
Starred rows report published NL2POSTCOND results~\cite{endres2024can} (840 tasks, 525 bugs); all others use our benchmark (1,134 tasks, 726 bugs).
The starred rows and our GPT-4o reconstruction use the condition-only format; all others use the four-field schema.

\smallskip\noindent\textbf{Comparable to GPT-4o, with NL2POSTCOND as a baseline.}
\specprobe{} achieves Sem@1 of 43.0\%, close to GPT-4o's 45.3\% (95\% CI: [40.1\%, 45.9\%] vs.\ [42.4\%, 48.2\%], $N{=}1{,}134$), BDR of 11.0\% vs.\ 12.1\%, and Exec@10 of 96.6\% vs.\ 98.5\%, at 141$\times$ lower inference cost (\$0.165${\times}10^{-3}$ vs.\ \$0.023 per generated postcondition).
Because NL2POSTCOND neither identifies which Defects4J bugs its results cover nor releases its Defects4J application code, we reconstructed its context-collection and assertion-insertion procedures from the paper and available logs~\cite{endres2024can} and applied them to our 1,134 tasks; this GPT-4o reconstruction achieves higher Sem@1 and lower Exec@1 than \specprobe{} (Section~\ref{rq3:prompt}).

\smallskip\noindent\textbf{Narrower gaps to GPT-5.5 with a candidate pool at 713$\times$ lower cost.}
While GPT-5.5 leads on Sem@1 (71.7\% vs.\ 43.0\%) and BDR (32.1\% vs.\ 11.0\%), the gap narrows with a candidate pool: \specprobe{} reaches 80.5\% vs.\ 88.4\% Sem@10, 94.7\% vs.\ 97.5\% Exec@5, and 96.6\% vs.\ 97.6\% Exec@10.
That is, from a top-10 pool, \specprobe{} yields at least one compilable, semantically consistent probe for 80.5\% of methods (vs.\ 88.4\%) at 713$\times$ lower inference cost (\$0.165${\times}10^{-3}$ vs.\ \$0.117), supporting the feasibility of incorporating it into CI workflows.

\smallskip\noindent\textbf{Best-in-class among open-source models, at the smallest size.}
Despite being the smallest model in the open-source group (7B, tied only with its unfine-tuned base), \specprobe{} ranks first on 5 of 8 performance metrics (all Sem@k, Exec@5, and Exec@10), outperforming models up to 32B, and trails the leader by only 2.4 points on Exec@1 (72.3\% vs.\ 74.7\%) and 1.7 points on BDR (11.0\% vs.\ 12.7\%).
Over its base model (Qwen2.5-Coder-7B), it yields a 3.4$\times$ Sem@1 gain (43.0\% vs.\ 12.5\%) and doubles Exec@1 (72.3\% vs.\ 34.9\%).
Against the unfine-tuned Qwen2.5-Coder-32B (${\sim}$4.6$\times$ as many parameters), \specprobe{} leads on Sem@k by 7--10 points, reveals 3 more bugs, and achieves a statistically significantly higher Sem@1 (two-proportion z-test, $p{<}10^{-6}$).
The strongest open-source competitor, DeepSeek-R1-Distill-32B, has approximately 4.6$\times$ as many parameters as \specprobe{}'s 7B model and uses test-time scaling; \specprobe{} nonetheless surpasses it on every Sem@k and Exec@k metric at 8.6$\times$ lower inference cost, while reaching 87\% of its bug count (80 vs.\ 92).

\begin{findingbox}[Finding 1]
\specprobe{}'s 7B model achieves Sem@1 close to GPT-4o's at 141$\times$ lower inference cost and leads the open-source baselines on all Sem@k metrics.
\end{findingbox}

\begin{figure}
    \centering
    \footnotesize
    \includegraphics[
    width=\linewidth,
    ]{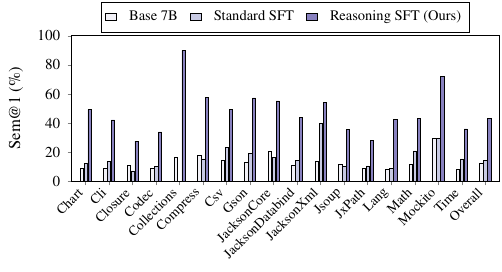}
    \caption{Sem@1: Base Qwen vs.\ Standard SFT vs.\ Reasoning SFT (Ours).
    }\label{fig:sft}
\end{figure}

\subsection{RQ2: Does reasoning justify its cost?}\label{sec:rq2}

Reasoning SFT adds training cost (curating reasoning traces) and inference cost (longer outputs).
We evaluate whether it improves probe quality, examining reasoning traces at training time (\autoref{fig:sft}) and reasoning length at inference time (\autoref{fig:trend}).

\subsubsection{Reasoning Traces vs.\ Standard SFT}

\autoref{fig:sft} compares Sem@1 for the base Qwen2.5-Coder-7B, a standard-SFT variant without reasoning traces, and \specprobe{}'s reasoning-SFT variant across all 17 Defects4J projects (1,134 tasks): 12.5\%, 14.2\%, and 43.0\% overall, respectively.
Standard SFT gains only 1.7 points over the base model, whereas reasoning SFT is 28.8 points above standard SFT and higher on all 17 projects.

\subsubsection{Reasoning Length}

\begin{figure}
    \centering
    \footnotesize
    \includegraphics[
    width=.45\textwidth,
    ]{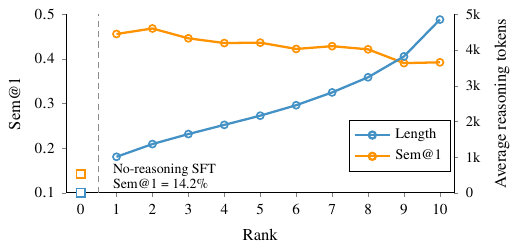}
    \caption{Sem@1 by Reasoning Token Length (Shortest to Longest). Rank~0 denotes the separate no-reasoning SFT baseline; ranks~1--10 order candidates from the reasoning model.
    }\label{fig:trend}
\end{figure}

\autoref{fig:trend} ranks each task's 10 candidates by reasoning-token count from shortest (rank~1) to longest (rank~10), and plots Sem@1 by rank, with the standard-SFT model as rank~0 (14.2\% Sem@1).
For reasoning SFT, Sem@1 reaches 46.8\% at rank~2 and decreases to 39.2\% at rank~10.
Reasoning thus matters, but longer is not necessarily better, consistent with prior findings on reasoning length~\cite{muennighoff2025s1}; length may therefore aid lightweight reranking (Section~\ref{sec:length-reranking}).

\begin{figure}[tb]
    \centering
    \footnotesize
    \includegraphics[
    width=.4\textwidth,
    trim=0cm 0.2cm 0cm 0cm,clip
    ]{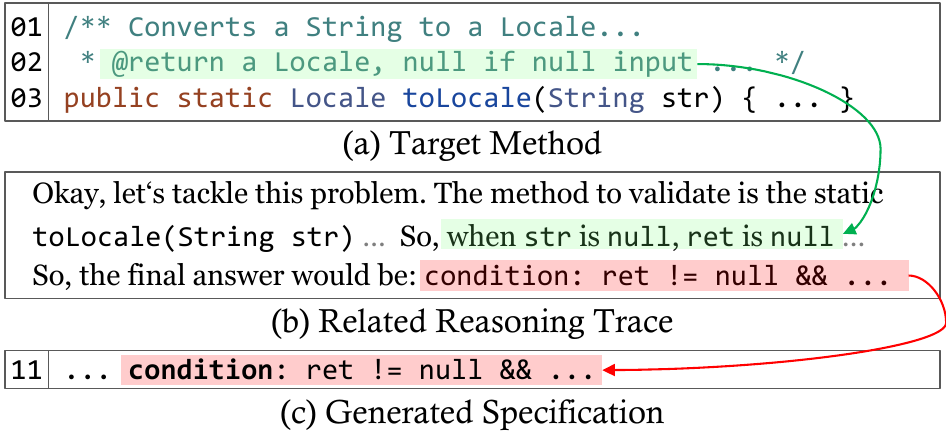}
    \caption{Reasoning Drift Example.}
    \label{fig:long_reasoning}
\end{figure}

\smallskip\noindent\textbf{Case study.}
\autoref{fig:long_reasoning} illustrates this failure mode.
For \code{toLocale(String str)} from \texttt{Lang}~bug-54, whose Javadoc explicitly specifies that a null input returns null, the model produces a 5{,}313-token reasoning trace.
Early on (token~1{,}524), it correctly states that ``when \code{str} is null, \code{ret} is null.''
After 3{,}393 more tokens, however, the final specification asserts \code{ret != null}, contradicting both the documentation and the model's own earlier reasoning.
This exemplifies \emph{reasoning drift}: prolonged reasoning gradually erodes a correct early conclusion, yielding a semantically inverted postcondition.

\begin{findingbox}[Finding 2]
Reasoning traces are essential for probe quality, but more is not always better: moderate-length reasoning maximizes semantic consistency, while excessively long traces degrade output through reasoning drift.
\end{findingbox}

\subsection{RQ3: What diagnostic evidence do probes provide beyond test stack traces?}

Defects4J supplies the methods modified by the fixes. We examine what diagnostic evidence probes provide once a target method has been selected.
A failing-test stack trace need not identify the method requiring a fix: the failure may surface in a test assertion, a downstream caller, or a framework path, hiding the root-cause method.
\specprobe{}'s postconditions flag observed violations at the target method, which may be absent from the stack trace.
Candidate-method selection is outside \specprobe{}'s scope; instead, \specprobe{} can be supplied with candidate methods by an existing debugging workflow.

\smallskip\noindent\textbf{Diagnostic evidence for methods absent from stack traces.}
Of the 80 bugs revealed by \specprobe{}, the native failing-test stack trace omitted the target method in 76 cases (95\%) and even its containing class in 73 (91\%).
For 307 of the 318 bug-revealing probes (96.5\%), the target method was absent from the original trace.

\begin{figure}[t]
    \centering
    \includegraphics[
    width=\linewidth,
    trim=0cm 0.2cm 0cm 0cm,clip
    ]{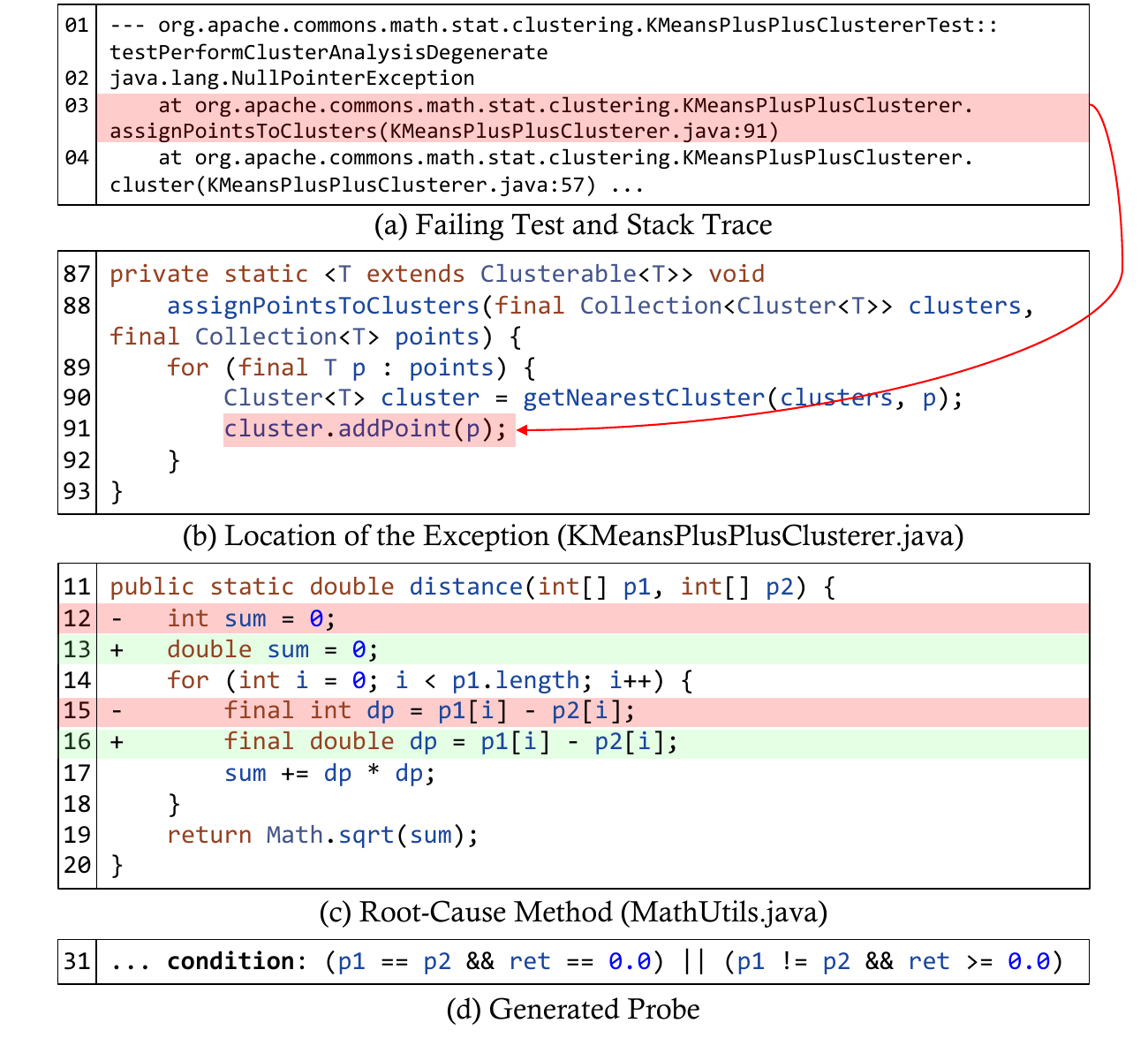}
    \caption{Bug-Revealing Probe for \texttt{Math} Bug-79.}
    \label{fig:bug_localization2}
\end{figure}

\smallskip\noindent\textbf{Case study: diagnostic evidence across files.}
\autoref{fig:bug_localization2} shows what a probe adds to the stack trace for \texttt{Math} Bug-79.
In the native failing-test log and stack trace (Figure~\ref{fig:bug_localization2}(a)), line~3 reports a \code{NullPointerException} at line 91 of \code{assignPointsToClusters()} in \code{KMeansPlusPlusClusterer.java} (Figure~\ref{fig:bug_localization2}(b)), the file the test directly exercises.
The true fault lies four call-chain hops away in a \emph{different file}: \code{distance()} in \code{MathUtils.java} (Figure~\ref{fig:bug_localization2}(c)) accumulates squared coordinate differences into an \code{int}, which silently overflows to a negative value; \code{Math.sqrt()} turns it into \code{NaN}, which propagates across file boundaries until it surfaces as the observed exception.
Neither the log, the stack trace, nor any symbol in \code{KMeansPlusPlusClusterer.java} references \code{distance()} or \code{MathUtils}, leaving a developer at the exception site with no textual foothold toward the root cause.
\specprobe{}'s probe (Figure~\ref{fig:bug_localization2}(d)) checks the non-negative invariant at the return site of \code{distance()}, catching the violation before \code{NaN} reaches the caller and reporting where and which condition fails.

\begin{findingbox}[Finding 3]
Bug-revealing probes report violated conditions at selected methods. For 95\% of revealed bugs, the original failing-test stack trace omits the probed method.
\end{findingbox}

\subsection{RQ4: Does the pipeline generalize across language, contract format, and curator model?}\label{sec:rq4}

\begin{table}[t]
\centering\footnotesize
\caption{Python Icontract Results on 164 HumanEval Tasks (bold: best per group). Ours uses GPT-5 for curation. MutD is evaluated at $k{=}10$.}
\label{tab:icontract}
\setlength{\tabcolsep}{3pt}
\begin{tabular}{lrrrr}
\toprule
Model/curation setting & Sem@1 & Sem@5 & Sem@10 & MutD \\
\midrule
GPT-5.5 & \textbf{83.8\%} & \textbf{90.4\%} & 92.1\% & \textbf{64.0\%} \\
GPT-5 & 56.2\% & 88.6\% & \textbf{92.7\%} & \textbf{64.0\%} \\
GPT-5-mini & 25.2\% & 62.5\% & 78.0\% & 49.4\% \\
GPT-o3 & 23.0\% & 59.6\% & 73.2\% & 47.6\% \\
GPT-4o & 11.1\% & 36.8\% & 53.0\% & 20.7\% \\
\midrule
Qwen2.5-Coder-7B & 4.4\% & 19.8\% & 35.4\% & 4.9\% \\
Qwen2.5-Coder-32B & 3.6\% & 15.3\% & 25.6\% & 8.5\% \\
DeepSeek-R1-Distill-32B & 10.5\% & 37.7\% & 53.7\% & 28.0\% \\
Standard SFT (3K) & 11.0\% & 32.8\% & 44.5\% & 29.9\% \\
Ours (1k) & 41.4\% & 70.2\% & 76.8\% & 51.8\% \\
Ours (2k) & 44.2\% & 74.5\% & 82.3\% & 53.0\% \\
Ours (3k) & \textbf{47.4}\% & \textbf{76.9}\% & \textbf{84.8}\% & \textbf{56.7}\% \\
\bottomrule
\end{tabular}
\end{table}

\begin{table}[t]
\centering\footnotesize
\caption{Curator Comparison on HumanEval Tasks. Ten candidates per task; MutD is evaluated at $k{=}10$.}
\label{tab:curator_compare}
\setlength{\tabcolsep}{3pt}
\begin{tabular}{lrrrr}
\toprule
Model/curator & Sem@1 & Sem@5 & Sem@10 & MutD \\
\midrule
Ours & 47.4\% & 76.9\% & 84.8\% & 56.7\% \\
Ours/DeepSeek-R1 & 58.9\% & 85.1\% & 91.5\% & 62.2\% \\
\bottomrule
\end{tabular}
\end{table}

We assess whether the pipeline's core mechanism (reasoning SFT with automated test-based filtering) transfers beyond Java (\autoref{tab:icontract}).
Unlike Java, which requires our custom injection schema, Python's icontract library~\cite{icontract} natively supports pre-state snapshots~(\texttt{old}) and postcondition decorators, so generated contracts execute without additional instrumentation.

\smallskip\noindent\textbf{Setup.}
We build training data from MBPP (974 methods)~\cite{austin2021program} with GPT-5 as the curator, yielding 3K prompt--reasoning--postcondition triples after automated test-execution filtering, and evaluate on HumanEval (164 methods)~\cite{liu2023your}.
We compare against GPT-5.5, GPT-5, GPT-5-mini, GPT-o3, and GPT-4o as proprietary references, alongside open-source baselines; no prior postcondition-generation baseline exists for this benchmark.
We report Sem@k (as in Java) and mutant-distinguishing rate (MutD), adopting NL2POSTCOND's mutants~\cite{endres2024can}, semantically incorrect variants of each HumanEval task, as ground-truth buggy versions.
MutD is the fraction of tasks for which a generated contract passes on the reference implementation and fails on all mutants, directly measuring how well it captures the method's intended semantics.
We also compare GPT-5 and DeepSeek-R1 as Python curators, holding the language, contract format, evaluation tasks, and base model fixed.

\smallskip\noindent\textbf{Higher Sem@k and MutD than GPT-5-mini and GPT-o3.}
\specprobe{} achieves 47.4\% Sem@1, 84.8\% Sem@10, and 56.7\% MutD with fully automated curation and only 3K training instances, far above the base Qwen2.5-Coder (4.4\% Sem@1, 4.9\% MutD).
It surpasses GPT-5-mini (25.2\% Sem@1, 49.4\% MutD) and GPT-o3 (23.0\% Sem@1, 47.6\% MutD) on Sem@k and MutD. GPT-5 achieves 56.2\% Sem@1 and 64.0\% MutD, and GPT-5.5 reaches 83.8\% Sem@1.
Qwen2.5-Coder-32B's low Sem@k stems mainly from icontract API misuse (e.g., unsupported decorator arguments) in 1,447 of \mbox{1,640} probes.
When DeepSeek-R1, the Java curator, supplies its own filtered Python training examples, the model achieves 58.9\% Sem@1, 91.5\% Sem@10, and 62.2\% MutD (\autoref{tab:curator_compare}), vs.\ 47.4\%, 84.8\%, and 56.7\% for the GPT-5-curated 3K model.

\smallskip\noindent\textbf{Reasoning traces remain essential; performance scales with training data.}
A non-reasoning SFT variant trained on the same GPT-5-curated data without reasoning traces degrades sharply to 11.0\% Sem@1 and 29.9\% MutD.
Increasing the training set from 1K to 3K yields consistent gains on both metrics (Sem@1: 41.4\% $\rightarrow$ 47.4\%; MutD: 51.8\% $\rightarrow$ 56.7\%).
As MBPP's 974 training methods yield ${\sim}$3K filtered triples, 3K is the practical ceiling here; we expect further gains with larger curated datasets.
Thus, the pipeline transfers to Python/icontract and remains effective with GPT-5 or DeepSeek-R1 curation.

\begin{findingbox}[Finding 4]
The pipeline transfers to a different language and contract format, and remains effective across the two evaluated curator models.
\end{findingbox}

\subsection{Data Contamination Analysis}\label{sec:training-overlap}

We conducted two complementary analyses to address potential data contamination.

\subsubsection{Training--evaluation overlap}
First, we audited overlap between the GitHub fine-tuning data and Defects4J. Following StarCoder~\cite{li2023starcoder} and LiCoEval~\cite{xu2025licoeval}, we computed Jaccard similarity over consecutive five-token sequences of target method declarations and comments. LiCoEval's 0.20 inspection threshold yielded 50 evaluation tasks (the 50th scored 0.2045; the 51st, 0.1915). Manually inspecting them and their closest training matches, we found potential semantic overlap only for the top three tasks (similarity 0.60 each); from the fourth-ranked task (0.44) onward, none was clear.

We conservatively removed all seven training examples of these three tasks and retrained from the same base model. The retrained model achieves 44.3\% Sem@1 and reveals 86 bugs (vs.\ 43.0\% and 80), so performance does not decrease and our main conclusions are not driven by these overlaps.

\subsubsection{Recent bug fixes}\label{sec:recent-bug-fixes}

Second, we evaluated the original model on 41 Java methods from 29 SWE-bench-Live~\cite{zhang2025swebenchgoeslive} bug-fix pull requests opened on or after July 1, 2025, after the base model's release and our fine-tuning dataset's creation.
With the same applicability criteria as for Defects4J, \specprobe{} achieves 44.4\% Sem@1 and 69.0\% Exec@1 (vs.\ 43.0\% and 72.3\% on Defects4J) and reveals 3 of the 29 bugs (\autoref{tab:swe-bench-live}), evidence that it remains effective on documented methods from recent real-world bug fixes with lower contamination risk.

\begin{table}[t]
\centering\footnotesize
\caption{Probe generation on 41 Java methods from 29 SWE-bench-Live bug fixes.}
\label{tab:swe-bench-live}
\setlength{\tabcolsep}{3pt}
\begin{tabular}{lrrr}
\toprule
Model & Sem@1 & Exec@1 & Bug-revealing rate \\
\midrule
GPT-5.5 & 82.4\% & 96.3\% & 13.8\% (4/29) \\
GPT-4o & 43.2\% & 76.1\% & 13.8\% (4/29) \\
\midrule
Qwen2.5-Coder 7B & 9.0\% & 29.8\% & 0.0\% (0/29) \\
Qwen2.5-Coder 32B & 30.5\% & 66.6\% & 6.9\% (2/29) \\
DeepSeek-R1-Distill & 43.9\% & 64.4\% & 13.8\% (4/29) \\
\specprobe{} & 44.4\% & 69.0\% & 10.3\% (3/29) \\
\bottomrule
\end{tabular}
\end{table}

\subsection{Analysis}

\subsubsection{Code Context Ablation}\label{sec:ablation}

We evaluate two ablated \specprobe{} variants across all 17 Defects4J projects (1,134 tasks), with training and inference prompts omitting repository-level context alone or both repository- and class-level context.
Sem@1 decreases from 43.0\% with full context to 41.6\% without repository context and to 28.4\% when class context is also removed (\autoref{tab:few_shot}): class-level context provides most of the gain, and repository-level context a smaller additional benefit.

\subsubsection{Few-Shot Prompting}

Prompting Qwen2.5-Coder-7B with 20 probe examples from outside Defects4J improves Sem@1 (12.5\%$\to$17.2\%) and Exec@1 (34.9\%$\to$41.5\%) (\autoref{tab:few_shot}); both Qwen baselines use the four-field schema.
This modest gain over zero-shot prompting does not close the gap to \specprobe{}, supporting the benefit of task-specific fine-tuning beyond schema prompting and in-context examples.

\begin{table}[!ht]
\centering\footnotesize
\caption{Few-Shot Prompting and Code Context Ablation on 1,134 Tasks. Results are for one candidate per task. Bug counts have denominator 726.}
\label{tab:few_shot}
\setlength{\tabcolsep}{3pt}
\begin{tabular}{lrrr}
\toprule
7B setting & Sem@1 & Exec@1 & Bugs ($k{=}1$) \\
\midrule
Qwen, zero-shot & 12.5\% & 34.9\% & 6 \\
Qwen, 20-shot & 17.2\% & 41.5\% & 8 \\
Wo. repo. \& class context & 28.4\% & 43.9\% & 10 \\
Wo. repo. context & 41.6\% & 70.6\% & 26 \\
\specprobe{} & \textbf{43.0\%} & \textbf{72.3\%} & \textbf{31} \\
\bottomrule
\end{tabular}
\end{table}

\subsubsection{False Alarms on Randomly Selected Methods}

\begin{figure}[t]
\centering
\includegraphics[width=0.9\columnwidth]{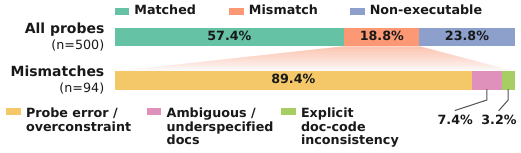}
\caption{Alerts on 500 Random Methods from 100 Defects4J Checkouts.}
\label{fig:false_alarms}
\end{figure}

We randomly sampled 100 Defects4J bugs and, excluding all methods modified by each fix (per the bug-fixing diff), randomly selected five other methods from each fixed version, yielding 500 probes: 94 (18.8\%) mismatched, 287 matched, and 119 (23.8\%) were non-executable (\autoref{fig:false_alarms}). Manual analysis of all 94 mismatches found 84 (16.8\% of all probes) spurious alerts caused by incorrect or overly restrictive constraints. The remaining ten (2.0\%) were potential documentation--implementation issues requiring human interpretation: seven involved ambiguous or underspecified documentation, and three were clear documentation--code inconsistencies. This quantifies the triage burden (18.8\% mismatched, 16.8\% confirmed spurious) and supports using \specprobe{} on selected candidate methods rather than indiscriminately repository-wide.

\subsubsection{Specification Schema}\label{rq3:prompt}

To isolate the effect of our four-field schema, we use GPT-4o with both output formats on the same 1,134 tasks (\autoref{tab:model_performance}). Compared with the NL2POSTCOND format, our schema yields slightly lower Sem@1 (45.3\% vs.\ 46.5\%) but substantially higher Exec@1 (85.4\% vs.\ 70.1\%) and more revealed bugs (88 vs.\ 57).
Of the 318 bug-revealing probes generated by \specprobe{}, 109 (34.3\%) use \texttt{pre-ghost} or \texttt{post-ghost}, fields that let probes compare pre- and post-call state when the pre-call value would be unavailable to a condition-only probe (\autoref{fig:moti}).

\subsubsection{Tautological Probes}
\begin{table}[t]
\centering\footnotesize
\caption{Sensitivity to Tautological Conditions. The filtered results remove 254 detected tautological or trivial probes from the same audited candidate pool; both settings retain the original ten-generation budget.}
\label{tab:triviality}
\setlength{\tabcolsep}{3pt}
\begin{tabular}{lrrrr}
\toprule
Setting & Sem@1 & Exec@1 & \shortstack{Bug-rev. probes} & Revealed bugs \\
\midrule
Original & 43.0\% & 72.3\% & 318 & 80 \\
Trivial-filtered & 40.7\% & 70.1\% & 318 & 80 \\
\bottomrule
\end{tabular}
\end{table}

Of all 4,874 matched \specprobe{} probes, we identified 254 (5.2\%) as tautological or trivially satisfied.
Excluding them lowers Sem@1 from 43.0\% to 40.7\% and Sem@10 from 80.5\% to 78.0\%, while all 318 bug-revealing probes and 80 revealed bugs remain unchanged (\autoref{tab:triviality}).
Such probes thus modestly inflate the semantic-generation metrics but do not affect our bug-revealing results.

\subsubsection{Cost and CI Overhead}
Java training-data curation with DeepSeek-R1 costs \$80.65, and fine-tuning takes eight A100 GPU-hours. As generation used concurrent requests, we did not record per-request latency; from observed output lengths and the reported per-token latency~\cite{qwenSpeedBenchmark}, we estimate 32.5~s per probe on one A100 80GB GPU at batch size one. Applying a probe takes 0.080~ms for insertion and 0.323~s for compilation and installation, and raises the relevant-test runtime from 6.045~s to 6.076~s (+0.031~s, 0.5\%); in total, probe application adds about 0.35~s per probe beyond the original test runtime.

\subsubsection{Error Analysis}
We automatically classify all 6,466 failed postconditions by their error logs into compilation failures (48.5\%), semantic overreach (46.2\%), and other runtime errors (5.3\%), and manually inspect 70 random cases from each of the two dominant categories (94.7\% of failures).
\textbf{Compilation failures} stem primarily from hallucinated API references: in 80\% of inspected cases, the model uses methods or fields absent from the target scope (e.g., \code{min}/\code{max} instead of the available \code{getMin()}/\code{getMax()}).
\textbf{Semantic overreach} yields compilable but over-constrained postconditions asserting stronger properties than the method guarantees, most commonly (72.9\% of inspected cases) by asserting exact equality on return values where only a weaker relational property holds.
The remaining runtime failures stem predominantly from \texttt{NullPointerException} on unguarded nullable state.
Compilation failures trace to weak grounding in repository context, consistent with the context ablation; semantic overreach reflects a generation-side bias that context alone cannot resolve: toward the strongest checkable property rather than the weakest sufficient one.

\subsubsection{Bug Analysis}
Reveal rate varies substantially across bug types, from 0\% for I/O stream bugs to 37.8\% for return-value correction bugs; a bug can belong to more than one category.

\smallskip\noindent\textbf{Return-value correction bugs.}
Bugs whose fixed patch adds exactly one effective \code{return} statement are revealed at 14/37 (37.8\%), 3.4$\times$ the overall rate.
These fixes concentrate the error in the returned value, making the violation directly observable by a postcondition.

\smallskip\noindent\textbf{Numerical computation bugs.}
Computational-logic bugs, exemplified by the \texttt{Math} project, are revealed at 28/100 (28.0\%), 2.5$\times$ the overall rate.
These faults lie in upstream arithmetic or algorithmic operations (e.g., integer overflow) whose errors largely propagate to the return value, where a postcondition can observe them.

\smallskip\noindent\textbf{Branch/loop modification bugs.}
Bugs whose patch introduces or modifies branching or loop structure are revealed at 42/515~(8.2\%), below the overall rate.
Such edits introduce more path-dependent behavior, making the faulty outcome harder for a postcondition to isolate and observe consistently across test executions.

\smallskip\noindent\textbf{I/O stream bugs.}
Bugs involving I/O stream operations are revealed at 0/39.
Their violated contracts are typically implicit or undocumented, leaving documentation-grounded probes no basis for the relevant check.
For \texttt{Codec} Bug-2, \texttt{encode} appends a line separator for empty input, violating the undocumented implicit contract that empty input yields empty output.

Together, these categories identify two determinants of reveal rate: whether the violated contract is documented and expressible as an executable postcondition, and whether the fault's effect is observable at the method's return site.

\subsubsection{Selecting probes by reasoning length}\label{sec:length-reranking}
We test reasoning length as a practical probe-selection signal.
We evaluated all 45 generation-and-selection configurations with $1 \le m < n \le 10$, where $n$ is the number of probes generated per task and $m$ is the number retained.
On average, retaining the shortest reasoning traces achieves 44.8\% Sem@1, vs.\ 43.0\% for random selection and 41.1\% for the longest.
For example, retaining the two shortest of ten yields 46.2\% Sem@1 and 75.1\% Exec@1, vs.\ 43.0\% and 72.3\% under random selection.
Reasoning length thus offers a simple inference-time signal for filtering probes toward higher-quality candidates.

\subsubsection{Revising non-executable probes}

\begin{figure}[t]
\centering
\includegraphics[width=0.8\columnwidth]{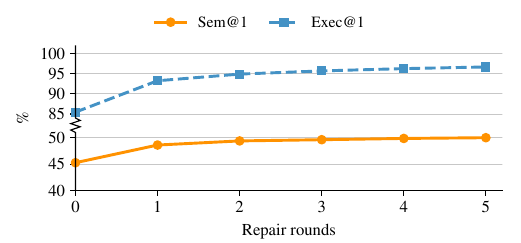}
\caption{Sem@1 and Exec@1 over zero to five rounds of probe revision.}
\label{fig:repair-rounds}
\end{figure}

We evaluate execution-feedback repair on 11,340 GPT-4o-generated probes (\autoref{fig:repair-rounds}). The first round increases Exec@1 from 85.45\% to 93.24\% and Sem@1 from 45.27\% to 48.62\%. After two rounds, Exec@1 reaches 94.89\% and Sem@1 49.40\%. Both metrics improve more slowly in rounds 3--5, reaching 96.66\% Exec@1 and 50.01\% Sem@1. Thus, repair is effective, with most of the benefit in the first two rounds.

\section{Related Work}\label{sec:related}

\noindent\textbf{Specification inference.}
Prior paradigms include static analysis~\cite{Chen2016SupportingOC,shoham2008static,Flanagan2001HoudiniAA,costa2007bouncer,cousot2013automatic,seghir2013counterexample}, dynamic invariant detection~\cite{ernst2001dynamically,nimmer2002automatic,astorga2018preinfer,Hangal2002TrackingDS,Ernst2007TheDS,molina2023specfuzzer}, and repository mining~\cite{ramanathan2007static,nguyen2014mining,Si2020Code2InvAD,Ryan2019CLN2INVLL,Molina2019TrainingBC}.
Dynamic tools such as Daikon~\cite{ernst2001dynamically} and SpecFuzzer~\cite{molina2023specfuzzer} generalize execution traces into likely invariants grounded in observed behavior rather than documented intent.
SpecGen~\cite{ma2024specgen} generates postconditions directly from code with runtime feedback, assuming the implementation is correct, and operates on toy programs without repository-level context.
\specprobe{} complements both: it derives behavioral hypotheses from NL documentation and uses execution to check alignment, surfacing mismatches that neither trace- nor code-based generation can detect.

\noindent\textbf{NL-to-specification generation.}
Earlier work relied on template-based extraction~\cite{tan2007icomment,tan2011acomment,tan2012tcomment,pandita2012inferring,goffi2016automatic,zhou2017analyzing,blasi2018translating,zhong2009inferring,phan2017statistical} or search-based synthesis~\cite{zhai2020c2s}.
Recent LLM-based approaches~\cite{kreber2021generating,cosler2023nl2spec,endres2024can,he2024beyond,xie2025effective,pan2023data,fuggitti2023nl2ltl,yang2023harnessing,yang2023leandojo,ma2024specgen,li2025extracting} show promise but typically depend on large proprietary models, runtime feedback during generation, or condition-only formats that do not explicitly represent repository-level dependencies or pre-state references.
Fine-tuning small models for specification generation has been explored~\cite{cao2025informal}, but with much larger training sets (14K examples) and without reasoning traces.
\specprobe{} addresses these limitations through a structured four-field schema that enables ghost variables and repository-level dependency declarations, reasoning SFT on only $\sim$1.5K automatically curated examples that eliminates dependence on large proprietary models at inference time, and instrumentation-based checking that closes the loop and optionally uses execution feedback to repair non-executable probes.
Unlike prior NL-to-spec work, \specprobe{} emphasizes specifications' executable role as behavioral probes yielding alignment evidence rather than correctness claims.
SpecMind~\cite{le2026specmind} also checks generated postconditions through execution but refines candidates with semantic feedback over multiple turns, whereas \specprobe{} preserves semantic mismatches for developer triage, since using them as repair feedback could mask genuine documentation--implementation discrepancies.
Concurrent work Expecto~\cite{lee2026expecto} synthesizes Dafny-style specifications from NL intent and validates them as final oracles with an SMT solver against input--output examples; \specprobe{} instead checks generated postconditions dynamically as probes against existing test suites to surface documentation--code misalignment, rather than producing a standalone specification.

\noindent\textbf{Documentation-code inconsistency.}
Work on detecting comment-code disagreements~\cite{tan2007icomment,tan2012tcomment,zhong2013detecting,zhou2017analyzing,panthaplackel2021deep,xie2022docter} motivates our conservative treatment of mismatches.
\specprobe{} differs in that the checked artifact is a model-generated executable postcondition, so the checker can report disagreement.

\section{Threats to Validity}\label{sec:threats}

\noindent\textbf{Internal validity.}
The primary threat is the fine-tuned vs.\ zero-shot comparison with proprietary baselines.
The RQ2 ablation mitigates it: with the same base model, standard SFT yields substantially lower Sem@1 than reasoning SFT, providing evidence that reasoning-trace supervision contributes to the improvement.
RQ4 separately shows that the same training-and-filtering pipeline applies to Python/icontract.
The Java fine-tuning and Defects4J repositories are disjoint. Our overlap audit and retraining assess possible overlap with the fine-tuning data (Section~\ref{sec:training-overlap}). Exposure to older benchmarks during base-model pretraining remains uncertain. Our fixed-base-model ablations hold pretraining exposure constant when assessing the effects of reasoning and code context. We also evaluated on fixes opened after the base-model release and the creation of the fine-tuning datasets (Section~\ref{sec:recent-bug-fixes}).

\noindent\textbf{Construct validity.}
Sem@k measures test-observed consistency, not full semantic correctness.
BDR mitigates this: a trivially true postcondition cannot distinguish fixed from buggy versions and therefore cannot inflate BDR.

\noindent\textbf{External validity.}
Defects4J features documented methods with test suites. The SWE-bench-Live evaluation covers 29 recent Java bugs whose documentation supports probe formulation; its smaller sample limits comparison precision with Defects4J. \specprobe{} assumes sufficient documentation and test coverage to generate and evaluate probes; poorly documented or weakly tested methods are outside its current scope.

\noindent\textbf{Reliability.}
All models use fixed decoding parameters; we sample 10 candidates per task and report 95\% confidence intervals for all key metrics.
All code, data, and generated probes are publicly released to support reproducibility.

\section{Conclusion}\label{sec:conclusion}

We presented \specprobe{}, a pipeline that generates executable behavioral probes for intent--implementation alignment assessment.
A structured schema, repository-aware prompts, and reasoning SFT on $\sim$1.5K examples enable a 7B model to achieve Sem@1 close to GPT-4o's at 141$\times$ lower cost and reveal 80 Defects4J bugs. The pipeline also transfers to Python, surpassing GPT-5-mini and GPT-o3 without human annotation. The probes report violated conditions at selected methods, complementing the original failing-test stack traces.

\section{Data Availability}\label{sec:data}
Following the Open Science Policy, we release source code, data, generated probes, and model weights for reproducibility at~\url{https://anonymous.4open.science/r/SpecProbe/}.

\bibliographystyle{IEEEtran}
\bibliography{spec,references}

\end{document}